\documentclass[10pt]{article}
\usepackage[OE]{express}

\usepackage{graphicx,color}
\usepackage{dcolumn}
\usepackage{bm}
\usepackage{epstopdf}
\usepackage{verbatim}
\usepackage{enumerate}
\usepackage{textcomp}

\usepackage[official]{eurosym}
\usepackage[squaren]{SIunits}
\usepackage{xcolor}
\usepackage[bookmarks=false]{hyperref}
\usepackage{color}
\usepackage{amsmath}

 \newcommand{\st}{\sigma_{\mathrm{t}}}
 \newcommand{\pmm}{p(\mu,\mu^\prime )}

\newcommand{\flm}{\psi_{\mathrm{lm}}(\textbf{r})}
\newcommand{\Ylm}{Y_{\mathrm{lm}}(\hat{\textbf{s}})}

\newcommand{\fm}{\psi_{\mathrm{m}}(z)} 
\newcommand{\fmpz}{\psi_{\mathrm{m+1}}(z)} 
\newcommand{\fmmz}{\psi_{\mathrm{m-1}}(z)}

\newcommand{\Izmud}{I_{\mathrm{d}}(z,\mu)}
\newcommand{\Izmupd}{I_{\mathrm{d}}(z,\mu^\prime)}

\newcommand{\pss}{p(\hat{\textbf{s}},\hat{\textbf{s}}^\prime )}

\newcommand{\Ibp}{I_{\mathrm{b}}(\textbf{r},\hat{\textbf{s}}^\prime)}

\newcommand{\Idd}{I_{\mathrm{d}}(\textbf{r},\hat{\textbf{s}})}

\newcommand{\stot}{\sigma_{\mathrm{t}}}

\newcommand{\sss}{\sigma_{\mathrm{s}}}
\newcommand{\sa}{\sigma_{\mathrm{a}}}

\newcommand{\Fo}{F_{\mathrm{0}}}

\newcommand{\Tbal}{T_{\mathrm{b}}(\lambda,L)}

\newcommand{\ls}{l_{\mathrm{s}}}
\newcommand{\lt}{l_{\mathrm{tr}}}
\newcommand{\la}{l_{\mathrm{a}}}

\newcommand{\Tt}{T(\lambda,L, \rho)}
\newcommand{\Rt}{R(\lambda,0, \rho)}

\newcommand{\IdT}{I_{\mathrm{d}}(\lambda, \mathrm{L}, \rho)}
\newcommand{\IdR}{I_{\mathrm{d}}(\lambda, \mathrm{0}, \rho)}
\newcommand{\Iio}{I_{\mathrm{i}}(\lambda, \mathrm{0}, \rho)}
\newcommand{\Iil}{I_{\mathrm{i}}(\lambda,\mathrm{L}, \rho)}

\newcommand{\Y}{\textrm{YAG:Ce}^{3+}}

\newcommand{\lai}{\lambda_{\mathrm{i}}}

\newcommand{\lac}{\lambda_{\mathrm{c}}}

\newcommand{\Ntr}{N_{\mathrm{tr}}}
\newcommand{\Nref}{N_{\mathrm{ref}}}
\newcommand{\Nabs}{N_{\mathrm{abs}}}
\newcommand{\Nb}{N_{\mathrm{b}}}

\newcommand{\FIGesmall}[5]{
	\begin{figure}[h!]
		\centering
		\includegraphics[width=#3]{#2.eps}
		\caption[#4]{\textbf{#4} #5}
		\label{#1}
	\end{figure}
}

\begin{document}

\title{Analytical modeling of light transport in scattering materials with strong absorption}

\author{M. L. Meretska\authormark{1,*}, R. Uppu\authormark{1}, G. Vissenberg\authormark{2}, A. Lagendijk\authormark{1}, W. L. IJzerman\authormark{2,3}, W. L. Vos\authormark{1} }

\address{\authormark{1}Complex Photonic Systems (COPS), MESA+ Institute for Nanotechnology, University of Twente, P. O. Box 217, 7500 AE Enschede, The Netherlands\\
\authormark{2}Philips Lighting, High Tech Campus 7, 5656 AE Eindhoven, The Netherlands\\
\authormark{3}Department of Mathematics and Computer Science, Eindhoven University of Technology, 5600 MB Eindhoven, the Netherlands}
\email{\authormark{*}m.meretska@utwente.nl} 

\begin{abstract}
We have investigated the transport of light through slabs that both scatter and strongly absorb, a situation that occurs in diverse application fields ranging from biomedical optics, powder technology, to solid-state lighting.
In particular, we study the transport of light in the visible wavelength range between $420$ and $700$ nm through silicone plates filled with YAG:Ce$^{3+}$ phosphor particles, that even re-emit absorbed light at different wavelengths. 
We measure the total transmission, the total reflection, and the ballistic transmission of light through these plates. 
We obtain average single particle properties namely the scattering cross-section $\sss$, the absorption cross-section $\sa$, and the anisotropy factor $\mu$ using an analytical approach, namely the P3 approximation to the radiative transfer equation. 
We verify the extracted transport parameters using Monte-Carlo simulations of the light transport. 
Our approach fully describes the light propagation in phosphor diffuser plates that are used in white LEDs and that reveal a strong absorption ($L/\la > 1$) up to $L/\la = 4$, where $L$ is the slab thickness, $\la$ is the absorption mean free path. In contrast, the widely used diffusion theory fails to describe this parameter range. 
Our approach is a suitable analytical tool for industry, since it provides a fast yet accurate determination of key transport parameters, and since it introduces predictive power into the design process of white light emitting diodes. 
\end{abstract}


\date{02 July 2017, in preparation for Optics Express}


\section{\label{ch:introduction}Introduction}
\begin{figure}[tbp]
\includegraphics[width=4.5 in]{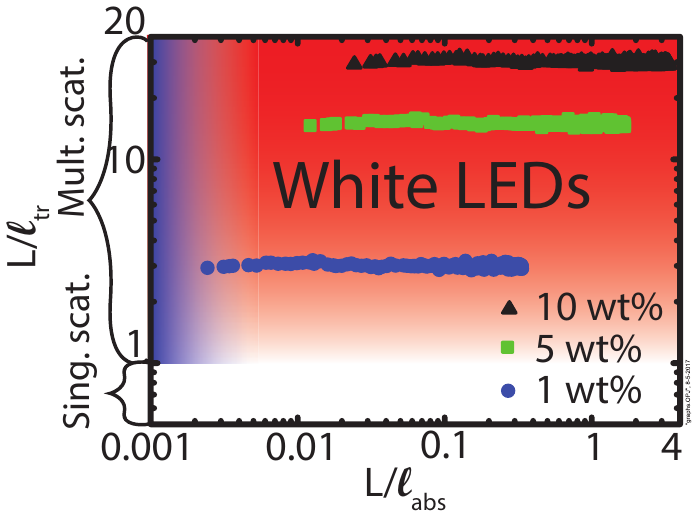}
\caption{\label{fig:cropped} 
\textbf{Transport parameters in the plane spanned by absorption and diffusion.}
The absorption on the abscissa is gauged by the ratio of sample thickness $L$ and absorption mean free path $\la$ and the diffusion strength on the ordinate by the ratio of sample thickness $L$ and transport mean free path $\lt$. 
The shaded blue range indicates the area where the diffusion theory performs well. 
The red range represents the transport parameter range accessed in the current manuscript. 
Symbols represent measured absorption and scattering as a function of the wavelength for phosphor diffuser plates with particle concentrations $1-10$ wt$\%$.}
\label{fig:parameterspace}
\end{figure}

Traditionally, optics has been concerned with clean or transparent components such as lenses, mirrors, beam splitters~\cite{Hecht1974Book}, whereas so-called 'dirty' components that scatter light were avoided as much as possible. 
Over the years, however, the realization has arisen that spatial inhomogeneities that scatter light have advantageous properties and allow applications that are otherwise impossible, for instance, an optical diffuser or a high-numerical aperture objective~\cite{ODonnell1987JOSAA,vanPutten2011PRL,Yilmaz2015Optica}. 
While both the know-how of and the control over optics that strongly scatters light has greatly advanced~\cite{Lagendijk1996PR,vanRossum1999RMP,Akkermans2007Book,Mosk2012NP,Wiersma2013NP,Rotter2017PhysRev}, the state-of-the-art is much less developed regarding optical systems that also strongly absorb light (or even re-emit light of a different color), even though important application fields occur in this regime, for instance solid-state lighting~\cite{Schubert2006Book,Krames2007JDT,Bechtel2008SPIE},  biomedical optics~\cite{Star1988PhysMedBio,Star1989SPIE,Pickering1993ApplOpt,Klose2006JComputPhys,Dickey1998PhysMedBiol,Dickey2001PhysMedBiol}, or powder technology~\cite{Burger1997ApplSpectrosc,Sekulic1996AnalChem,Shinde1999JPharmSci}. 

We illustrate the relevant parameter space of scattering and absorbing optical systems in Figure~\ref{fig:parameterspace}, where the abscissa represents the relative absorption strength and the ordinate represents the relative scattering strength.
The scattering strength is shown as the ratio of the sample thickness $L$ and the transport mean free path $\lt$.
The transport mean free path gauges how diffuse an incident light beam has become upon multiple scattering since it is equal to the average distance after which a directional incident light beam is randomized~\cite{Lagendijk1996PR,vanRossum1999RMP,Akkermans2007Book}. 
The absorption strength is shown as the ratio of $L$ and the absorption mean free path $\la$.
The absorption mean free path is the average diffuse propagation distance after which scattered light is absorbed to a fraction $(1/e)$. 

As an important practical class of a scattering and strongly absorbing optical system, we focus in this paper on white light emitting diodes (LEDs). 
A typical white LED consists of a blue semiconductor LED as a source and a phosphor layer~\cite{Schubert2006Book,Krames2007JDT,Bechtel2008SPIE}. 
In the phosphor layer, micrometer-sized phosphor particles made of, for instance, YAG:Ce$^{3+}$ scatter multiple times the light and absorb part of the light that is emitted by the blue LED. 
The absorbed blue light is re-emitted by the phosphor as a mixture of green, yellow and red light that together with remnant blue light produce the desired white light.
Furthermore, the multiple scattering of both the blue and the re-emitted light results in diffuse white light, which is crucial for an even illumination over a large area. 

Currently, the light propagation through scattering and strongly absorbing optical systems, such as white LEDs, is described using various numerical methods such as Monte Carlo simulations and ray tracing~\cite{Gilray1996IESNA,Cassarly2001Book,Sommer2009JSTQE,Liu2010AO,Tukker2010SPIE}.
While the methods are from the outset flexible to the detailed geometry of the problem at hand, they also have several limitations. 
There is a trade-off between the precision and the speed of numerical approaches that ultimately compromises the design of white LEDs, since precision increases with the number of numerical samples taken, thereby obviously decreasing speed. 
Moreover, the accuracy of numerical methods is not always beyond doubt, as \textit{a posteriori} readjustments of the transport parameters have been reported~\cite{Liu2010AO,Tukker2010SPIE}. 
Numerical methods do not possess predictive power, as opposed to analytical methods, hence for every new situation (\textit{e.g.}, new phosphor, new blue LED, etc.) new simulations must be performed. 
Therefore, it is timely to search for analytic approaches that have predictive power (even outside the immediately studied domain), and that are accurate and fast. 

In the weakly scattering regime $L/\lt << 1$ (see Fig.~\ref{fig:parameterspace}), the transport of light is analytically described by the venerable Lambert-Beer (or Beer-Lambert-Bouguer) law. 
In scattering systems such as white LEDs, however, the condition $L/\lt \approx 1$ holds~\cite{Vos2013AO,Leung2014OE}, hence light is scattered multiple times and diffused~\cite{footnote:ellscat-vs-elltr}. 
Moreover, we will see that in certain common circumstances, light is also strongly absorbed $L/\la \ge 1$. 
A widely known theory of transport of light is the radiative transfer equation~\cite{Chandrasekhar1960Book}. 
While the radiative transfer equation generally requires numerical solutions, there are several relevant cases where analytical solutions exist. 
The first order approximation to radiative transfer equation is the diffusion equation~\cite{Kaveh1991Book,Lagendijk1996PR,vanRossum1999RMP,Aydin2002IJMP}.
Solutions to the diffusion equation work surprisingly well even in regimes where they are not supposed to function~\cite{Lagendijk1996PR} such as optically thin samples~\cite{Vos2013AO}. 
Regarding absorption, however, the validity of the diffusion theory is limited to weakly absorbing systems with $L/\la\ll 1$, as indicated in Fig.~\ref{fig:parameterspace}~\cite{vanRossum1999RMP}. 
In white LEDs, however, the absorption mean free path at maximum absorption is at least four times smaller than the sample thickness $L/\la\geq 4$, see Fig.~\ref{fig:parameterspace}. 
Since in this regime the diffusion theory fails to describe the light propagation, other analytical approaches are required to accurately determine transport parameters of scattering materials with absorption, and to offer fast design tools for white LEDs.

In this paper we describe an analytical, accurate, and fast approach to determine the transport parameters in samples that are used in white LEDs. 
The determination of the transport parameters involves the measurement of the total transmission, total reflection, and the ballistic transmission for a set of diffuser plates as a function of the concentration of the phosphor particles. 
The total transmission and the total reflection data are analyzed using the P3 approximation to the radiative transfer equation (RTE)~\cite{Star1989SPIE}. 
In optically thin scattering samples where the contribution of the ballistic light dominates, the data were analyzed using the Lambert-Beer law. 
To verify our approach we performed Monte-Carlo simulations, and found excellent agreement between numerically obtained transport parameters and analytically obtained transport parameters.
The demonstrated approach offers a fast and accurate determination of the transport parameters and possesses predictive power beyond the parameter range studied. 
The precision of the extracted transport parameters is only limited by the statistical error of the experimental data. 
In addition, the proposed approach of measuring the total transmission $T$ and the total reflection $R$ allows us to decouple the scattering and the absorption in a straightforward experimental way. 
Once transport parameters are known they can be used as input parameters in the ray tracing Monte-Carlo calculations for white LED design which require complex geometries that are not captured by the analytical approach~\cite{LightTools2017}.  
We show that the parameter extraction is also applicable within the domain of validity of the diffusion theory. 
Hence, the proposed parameter extraction approach pertains also to the propagation of light in other relevant application fields, such as oceans and clouds~\cite{Funk1973ApplOpt,Zege1993ApplOpt,Fournier1994ProcSPIE,Fell2001JQuantSpectrosRadiatTransfer,Stramski2004ProgrOceanogr}, pharmaceutical products~\cite{Burger1997ApplSpectrosc,Sekulic1996AnalChem,Shinde1999JPharmSci}, and in noninvasive diagnostic imaging of living tissues~\cite{Star1989SPIE,Star1988PhysMedBio,Pickering1993ApplOpt,Klose2006JComputPhys,Dickey1998PhysMedBiol,Dickey2001PhysMedBiol}. 

\FIGesmall{fig:scat}{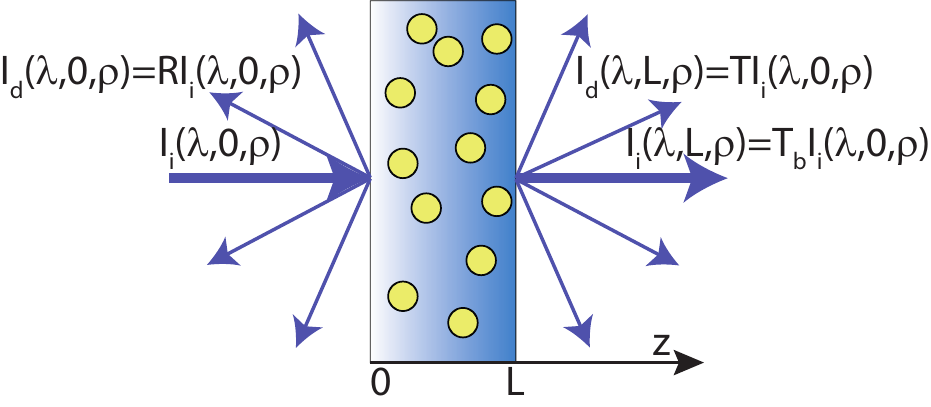}{3 in}{Scheme of light incident on and exiting from a slab.}
{Plane waves with intensity $\Iio$ are incident in the $z-$direction on a slab of scattering material that contains phosphor particles (represented by yellow circles). 
$\IdR$ is the diffuse reflected intensity, $\IdT$ is the diffuse transmitted intensity, and $\Iil$ is the transmitted ballistic intensity. 
$T$ and $R$ are the total transmission and the total reflection, respectively, and $T_\mathrm{b}$ is the ballistic transmission.}

\section{\label{sect:theory_meas}Model}
In this paper the diffuser plate with phosphor particles is modeled as a slab of thickness $\mathrm{L}$ that extends from $z = 0$ to $z = \mathrm{L}$, as shown in Fig.~\ref{fig:scat}.
Incident plane waves at wavelength $\lambda$ with intensity $I_{\mathrm{i}}(\lambda, z = 0, \rho) = I_{\mathrm{i}}(\lambda, 0, \rho)$ illuminate the slab. 
The incident light is scattered multiple times and possibly absorbed inside the slab.
We characterize the light scattering and absorption in the slab by measuring the total transmission $T$, the total reflection $R$, and the ballistic transmission $T_\mathrm{b}$ that are defined as
\begin{equation} 
\Tt \equiv \frac{\IdT+\Iil}{\Iio}
 \mathrm{,}
\label{eq:ri}
\end{equation}
and
\begin{equation} 
\Rt \equiv \frac{\IdR}{\Iio}
 \mathrm{,}
\label{eq:ri}
\end{equation}
and
\begin{equation} 
\Tbal \equiv \frac{\Iil}{\Iio}
 \mathrm{,}
\label{eq:ri}
\end{equation}
where $\IdR$ and $\IdT$ are the diffused intensity integrated over all outgoing angles at the positions $z = 0$ and $z = \mathrm{L}$, respectively, and $\Iil$ is the ballistic transmitted intensity. The measured total transmission T, the total reflection R, and the ballistic transmission $T_\mathrm{b}$ explicitly depend on the experimentally available parameters: the wavelength $\lambda$, the thickness of the sample $L$, and the density $\rho$ of the phosphor particles. We will use measured total transmission $T$, total reflection $R$, and ballistic transmission $T_\mathrm{b}$ to extract average single particle properties: the scattering cross section $\sss$, the absorption cross section $\sa$, and the anisotropy factor $\mu$~\cite{Bohren1983Book}.

To extract average single particle properties $(\sss,\sa,\mu)$ we exploit an analytical model to compute the total transmission $T^{\mathrm{m}}$, the total reflection $R^{\mathrm{m}}$ and the ballistic transmission $T^{\mathrm{m}}_\mathrm{b}$ that not only explicitly depend on the wavelength $\lambda$, the thickness of the sample $L$, and the density $\rho$ of the phosphor particles, but also on single particle properties $(\sss,\sa,\mu)$. By equating ($T^{\mathrm{m}}$,  $R^{\mathrm{m}}$, $T^{\mathrm{m}}_\mathrm{b}$) to the measured ($T$,  $R$, $T_\mathrm{b}$) we compute the transport parameters $(\sss,\sa,\mu)$. As the analytical model we chose to employ the P3 approximation to the radiative transfer equation~\cite{Star1989SPIE,Boas1995ProcSPIE,Dickey1998PhysMedBiol,Dickey2001PhysMedBiol,Faris2005ApplOpt,Klose2006JComputPhys,Liemert2014MedPhys}
\begin{equation}
T^{\mathrm{m}}(\lambda,L,\rho;\sss,\sa,\mu) \equiv \frac{F(\mathrm{L})}{\Fo}
\mathrm{,}
\label{eq:f2a}
\end{equation}
and
\begin{equation}
R^{\mathrm{m}}(\lambda,L,\rho;\sss,\sa,\mu)\equiv\frac{F(0)}{\Fo}
\mathrm{,}
\label{eq:f2b}
\end{equation}
with 
\begin{equation} 
F(z)=\sum\limits_{i=1}^{4}B_{\mathrm{1i}} \exp{(\mu_i z)}+G_1 \exp{(-\rho\stot z)} 
 \mathrm{,}
\label{eq:pt}
\end{equation}
where $\stot=\sss+\sa$ is the sum of the scattering cross section $\sss$ and the absorption cross section $\sa$, and $B_{\mathrm{1i}}$, $G_{\mathrm{1}}$ and $\mu_\mathrm{i}$ are functions described in the Appendix, $\rho$ is the particle concentration in the samples, and $\Fo$ is the incident light flux and $F(z)$ the flux of the light at position z.

The ballistic intensity that is transmitted through the sample is described by the Lambert-Beer law
\begin{eqnarray}
T^{\mathrm{m}}_\mathrm{b} (\lambda,L,\rho;\sa,\sss) \equiv 
\frac{\Fo e^{-(\rho\stot L)}}{\Fo}=\exp(-\rho\stot L)
 \mathrm{.}
\label{eq:bal1}
\end{eqnarray}
Knowledge of the average single particle scattering properties ($\sss$, $\sa$, $\mu$) at any wavelength $\lambda$ allows us to \textit{predict} the light transport properties for any scattering and absorbing sample for any particle density $\rho$ and sample thickness $L$ outside the immediately studied density and thickness range. 
Therefore, it is convenient to express the single-particle properties into the scattering mean free path $\ls$ the transport mean free path $\lt$ and the absorption mean free path $\la$ from the following relationships~\cite{Leung2014OE,Meretska2016ContractorRept}
\begin{equation} 
\ls=\frac{1}{\rho \sss}
 \mathrm{,}
\label{eq:pt1}
\end{equation}

\begin{equation} 
\la=\frac{1}{\rho \sa}
 \mathrm{,}
\label{eq:pt2}
\end{equation}

\begin{equation} 
\frac{1}{\lt}=\frac{(1-\mu)}{\ls}+\frac{1}{\la}
 \mathrm{,}
\label{eq:pt3}
\end{equation}
provided the underlying assumptions of the independent scatterer approach~\cite{Lagendijk1996PR} (negligible multiparticle effects, negligible re-absorption) remain fulfilled. 

\section{Experimental details}

The samples studied in the experiments are $L=1.98\pm0.02~\textrm{mm}$ thick silicone plates (polydimethylsiloxane, PDMS, refractive index $n = 1.4$~\cite{Querry1987ContractorRept}) that are doped with $\Y$ phosphor particles in a range of particle concentration $\rho$ ranging from $1$ to $10$ weight percent phosphor. 
The absorption spectrum of the phosphor particles has a peak absorption  wavelength $\lambda=458$~nm, and a broad linewidth (full width half maximum, FWHM) of $\Delta\lambda_a=55$~nm, as reported earlier in Ref.~\cite{Meretska2016ContractorRept}. 
Based on the absorption spectrum we choose the wavelength $\lac=520$~nm to distinguish the absorbing range ($\lambda=400$~nm to $520$~nm) from the non-absorbing range ($\lambda=520$~nm to $700$~nm). 

\FIGesmall{fig:set}{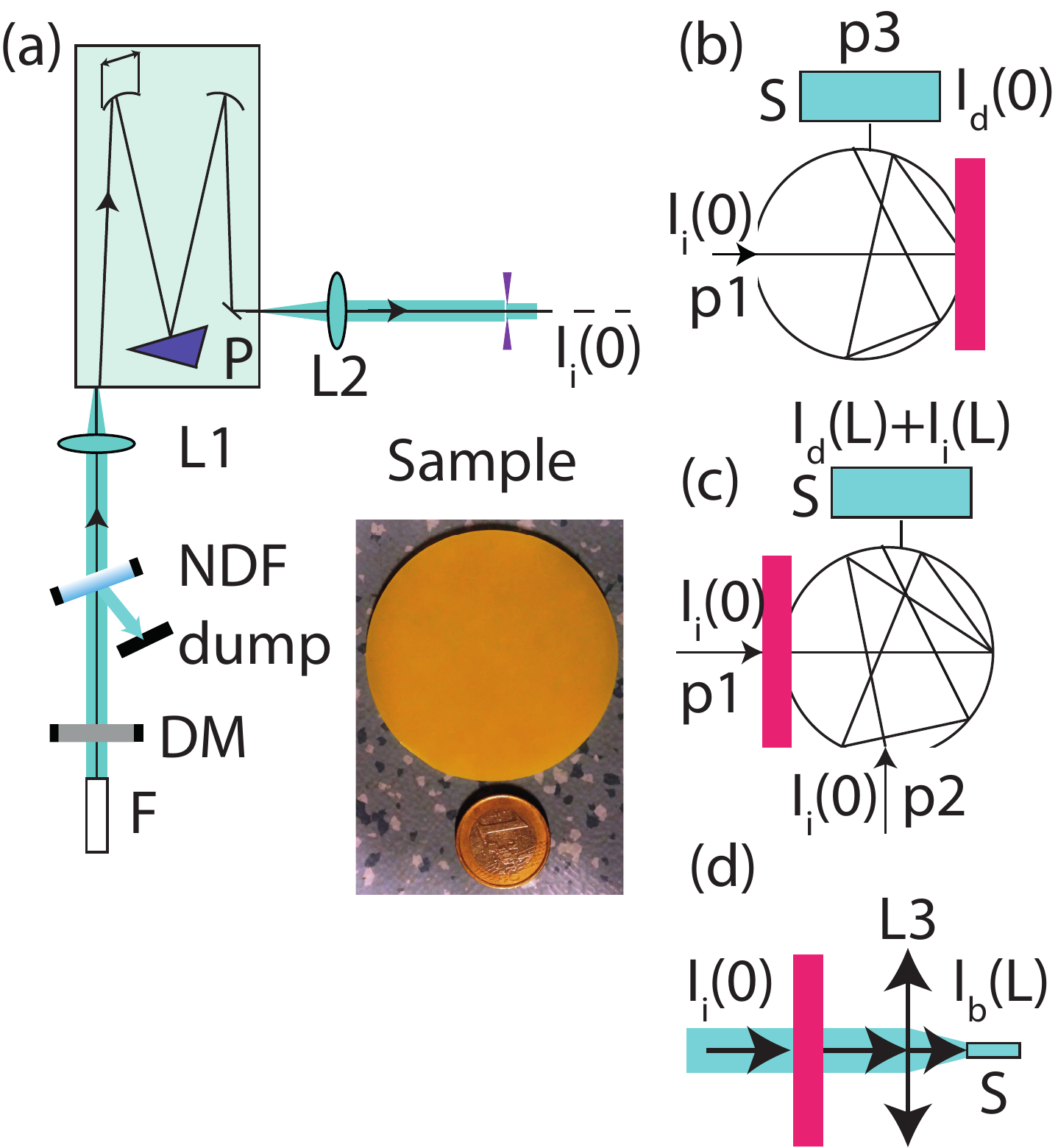}{3 in}{Experimental setup.}{
\textbf{(a)} The light source used in the experiment.
\textbf{(b)} The configuration of the integrating sphere used for the total reflection measurements. 
\textbf{(c)} The configuration of the integrating sphere used for the total transmission measurements. 
\textbf{(d)} The configuration for ballistic Lamber-Beer measurements.  
F: Supercontinuum white light source, NDF: Neutral density filter, DM:
Dichroic mirror, L1: Achromatic doublet (AC080-010-A-ML, f=10 mm), L2: Achromatic doublet (f=50 mm), L3: Achromatic doublet (AC254-050-A-ML, f=50mm), I: Integrating sphere, S: Spectrometer, P: Prism monochromator $(f_{\sharp}=4.6)$, p1: port1, p2: port 2, p3: port 3.  }

Figure~\ref{fig:set} shows a schematic of the experimental setup used to measure the total transmission $T$, the total reflectance R and the ballistic transmission $T_\mathrm{b}$. 
The sample is illuminated with a tunable narrowband light source in the visible wavelength range 420 -- 700 nm.
To this end, the beam of a supercontinuum white-light source (Fianium WL-SC-UV-3) is spectrally filtered to an adjustable bandwidth $\Delta\lambda<2.4$ nm using a prism monochromator (Carl-Leiss Berlin-Steglitz). 
Unwanted emission of the supercontinuum source in the infrared $\lai>$~700 nm is blocked with a neutral density filter (NENIR30A) and a dichroic mirror (DMSP805) (see Fig.~\ref{fig:set}(a)).

The incident beam illuminates the sample at normal incidence. 
The scattered intensity is collected using either an integrating sphere or a ballistic detector. 
The integrating sphere (Opsira uku-240) has three entrance ports, each with a $20$~mm diameter. 
Each port can be selectively closed with a baffle that has the same diffusive inner coating as the remainder of the integrating sphere. 
The entrance port of the integrating sphere is sufficiently large to collect scattered light that emanates at all angles from the strongest scattering sample. 
The intensity of the outgoing light entering the integrating sphere is analyzed with a fiber-to-chip spectrometer (AvaSpec-USB2-ULS2048L) with a spectral resolution $\Delta\lambda_\mathrm{s}=2.4$~nm. 

To measure the diffuse reflectance $R$, the ports 1 and 3 of the integrating sphere are used, see Fig.~\ref{fig:set}(b). 
The sample is attached to port 3 with the incident light intensity $\Iio$ entering the sphere through port 1. 
The diffuse reflected light intensity $\IdR$ is collected with the fiber spectrometer.

In the total reflection configuration, the reference intensity $\Iio$ separately enters the sphere under an angle through port 1, in such way that the beam is incident on the surface of the sphere but not onto the sample.
To measure the total transmission, the ports 1 and 2 are used, as shown in Fig.~\ref{fig:set}(c). 
The sample is attached to port 1 with the incident light intensity $\Iio$ illuminating the sample. 
The transmitted light is collected by the integrating sphere and spectrally resolved with the fiber spectrometer. In the transmission configuration, the reference intensity $\Iio$ is separately measured with the incident light entering sphere through port 2.

The ballistic detector measures the transmitted ballistic light $T_\mathrm{b}$. 
The scheme of the detector is shown in Fig.~\ref{fig:set}(d). It consists of an achromatic lens (f=50 mm) that collects the transmitted light into the entrance of a fiber spectrometer (AvaSpec-USB2-ULS2048L). 
The distance between the detector and the sample is set to $50$~cm.
If the detector is placed closer to the sample, the ballistic signal is compromised by a significant contribution from the diffuse scattered light. 
The contribution of the scattered light in the ballistic signal is estimated following Ref.~\cite{Li2016ApplOpt} to be less than $1\%$.

The measurements of the transmission $T$, the reflection $R$ and the ballistic transmission $T_\mathrm{b}$ are repeated 10 times at every wavelength, to obtain statistical information. 
We estimated the error bar to be about $\Delta \mathrm{T}/\mathrm{T}=\Delta \mathrm{R}/\mathrm{R}=4$~percent point for transmission and reflection measurements, and about $\Delta T_\mathrm{b}/ T_\mathrm{b}=1$~percent point for the ballistic transmission measurements. 
In Fig.~\ref{fig:sssa} and Fig.~\ref{fig:Tall} the error bars are within the symbol size. 
The estimated errors of the total transmission $T$, and the total reflection $R$ are propagating linearly when the transport parameters are calculated, and result in a 4 percent point error on the extracted transport parameters ($\sss$, $\sa$, $\mu$).

\section{Simulations}

We perform Monte-Carlo ray tracing simulations, in other words, Monte-Carlo simulations of the radiative transfer equation. The Monte-Carlo ray tracing was carried out using a computational cell similar to the schematic in Fig. 2 with normally incident monochromatic plane wave on an plane parallel slab with finite thickness $L$. 
The scattering parameters ($\ls$, $\la$, $\mu$) for the chosen wavelength are taken as the input parameters for the simulations. 
The light transport is simulated as follows~\cite{Muju2010JNP,Uppu2013PRA}: 
A bunch of $\Nb=10^4$ random walkers (photons) is launched with initial angular coordinates ($\theta_0$, $\phi _0$) = (0 rad, 0 rad) representing the normally incident plane wave. 
The incident photon bunch undergoes a three-dimensional random walk inside the slab. 
The random walk consists of a set of connected rectilinear paths of lengths \{$l_i$\}, each of which is randomly picked from an exponential distribution $p(l_i) = (1/\ls) \exp(-l_i/\ls)$. 
The scattering at the end of each path is simulated by choosing an angular coordinate ($\theta_i$, $\phi_i$) following the Henyey-Greenstein phase function to accommodate the scattering anisotropy $\mu$ of the scatterers. 
Absorption in the medium is incorporated through the exponential decrease of the photon number: the attenuation depends exponentially on the path length $l_i$ and on the absorption mean free path $\la$. 
The random walk is terminated when the photon bunch is either completely absorbed or it has arrived at an interface. 
The specular light reflection from the interface is described with Fresnel's law~\cite{Hecht1974Book}. 
The amount of transmitted photons $\Ntr$, reflected photons $\Nref$, or absorbed photon $\Nabs$ weight is registered at the end of the walk. Simulated transmission is defined as $T^{\mathrm{s}}\equiv \Ntr/\Nb$, and simulated reflection as $R^{\mathrm{s}}\equiv\Nref/\Nb$.
A total of about $N_{\mathrm{ph}}=2\times$10$^4$ photon bunches are launched to statistically compute the transmission and reflection for a given set of transport parameters ($\ls$, $\la$, $\mu$). The number of photons $N_{\mathrm{ph}}$ is chosen to ensure the convergence of the algorithm within 1$\%$ of the transmission fluctuations between different algorithm runs.
We scan the transport parameter range ($\mu \in  [0,1]$, $\sigma_s \in [1,3]$ and $\sigma_a \in [0,0.5]$) for different phosphor concentrations and choose the transport parameters that yield the transmission and the reflection obtained in the experiment.

\section{Results}
\subsection{Ballistic transmission}
The samples with 1~wt$\%$ and 2~wt$\%$ phosphor concentration were used for the ballistic transmission measurements since they are optically sufficiently thin. 
The ballistic transmission for other plates is less than 0.1$\%$, which is below the signal to noise level of our detector at the available incident signal intensities.

\FIGesmall{fig:bal}{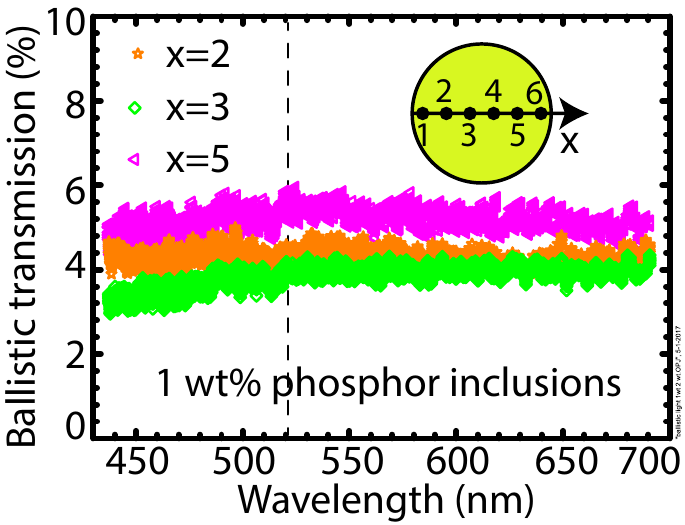}{3 in}{Ballistic transmission}{of the silicone plate with 1~wt$\%$ of phosphor particles measured at different spatial positions. 
The beam position during the measurements is shown in the inset.}

We scan the incident wavelength $\lai$ from 420~nm to 700~nm, and collect the ballistic light, with light incident at six different spatial positions on the sample as shown in the inset of Fig.~\ref{fig:bal}. 
The ballistic transmission spectra for the slab with 1~wt$\%$ is shown at three selected spatial positions in Fig.~\ref{fig:bal}. 
The ballistic transmission at wavelength $\lai$ differs at most by 2 percent points at different spatial locations $\Delta T_\mathrm{b}/T_\mathrm{b} < 2\%$. 
The difference between different data sets arises from the exponential sensitivity of the ballistic light to variations in the local phosphor particle concentration (see Eq.~\ref{eq:bal1}). 

\FIGesmall{fig:sssa}{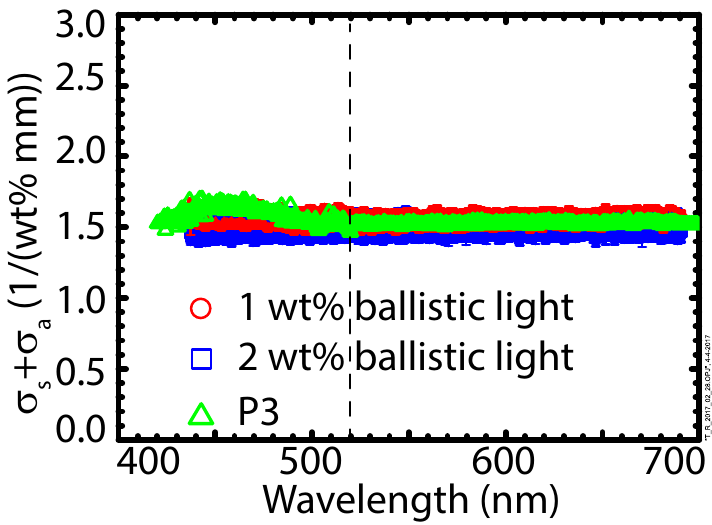}{3 in}{Scattering and absorption cross section.}{The sum of the scattering cross section $\sss$ and absorption cross section $\sa$ as a function of wavelength for two different phosphor concentrations. 
Green triangles represent data extracted from the diffused light measurements.}

We extract the sum of the scattering and absorption cross sections $\sss+\sa$ from the measured ballistic transmission data using Eq.~\ref{eq:bal1}. 
The results are plotted in Fig.~\ref{fig:sssa} for the samples with 1~wt$\%$ and 2~wt$\%$ concentrations of phosphor particles. 
We observe an excellent agreement between samples with different concentrations of phosphor in the wavelength range of interest. 
In the non-absorbing range (520 nm to 700 nm), the extracted cross section is equal to the scattering cross section $\sss$ as the absorption cross section $\sa\to 0$. 

In the absorption range the measured ballistic transmission is constant within the signal to noise ratio (SNR). 
From this behavior we conclude that the absorption cross section $\sa$ is at least one order of magnitude smaller than the scattering cross section $\sss$. 
In the case of the multiple scattering even this small absorption have crucial influence on the diffuse transmission, and reflection (see subsequent sections).

\subsection{Diffuse light transmission and reflection}

\FIGesmall{fig:Tall}{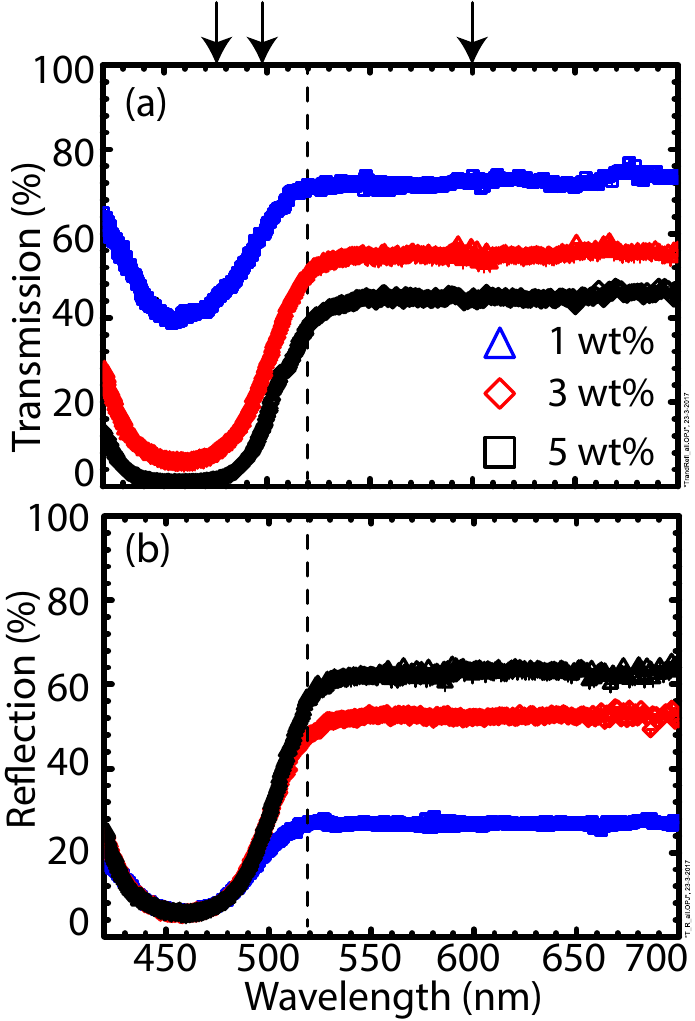}{3 in}{Transmission and total reflection spectra of the silicone plates}{\textbf{(a)} Broadband transmission $T$ as a function of concentration. \textbf{(b)} Broadband reflection as a function of concentration. Concentration of the phosphor particles is indicated on the figures. Arrows indicate the wavelengths for the Fig.~\ref{fig:Tw}}

We measure the transmission and the reflection spectra for the slabs with different phosphor particle concentrations. 
The measured results are shown in Fig.~\ref{fig:Tall}(a) and (b). 
For visual clarity, we plot only a selection of the measured samples. The transmission and the reflection spectra reveal a deep trough with a minimum at $\lai=458$~nm. 
The trough matches with the peak of the absorption band of $\Y$~\cite{Meretska2016ContractorRept}. 
The presence of the trough indicates that a significant fraction of the light in this wavelength range is absorbed by the phosphor. 
In the range $490~\mathrm{nm}<\lai<520~\mathrm{nm}$ where absorption and emission of phosphor overlap~\cite{Meretska2016ContractorRept}, the spectra show a steep rise in the transmission due to the reduced absorption. 
At wavelengths longer than $\lai=520$~nm, the transmission spectra are flat. Unlike the transmission spectra, the total reflection spectra depend only weakly on the phosphor concentration $\rho$ in the absorption range up to $\lambda=490$~nm (Fig.~\ref{fig:Tall}(b)).

The sum of the total transmission and the total reflection was found to be $104~\%$ for all the measurements in the non-absorbing range. 
This small systematic error could arise from an overestimation of the diffuse light signal. 
A possible source of the excess signal is the specular light (see Fig.\ref{fig:set}). 
Approximately $2~\%$ of the reference light is being specularly reflected (Fresnel reflection) from the sample, and is lost in the setup. 
Part of the specular reflected light can also illuminate the sample, however, resulting in an effectively higher intensity that illuminates the sample. 

\FIGesmall{fig:Tw}{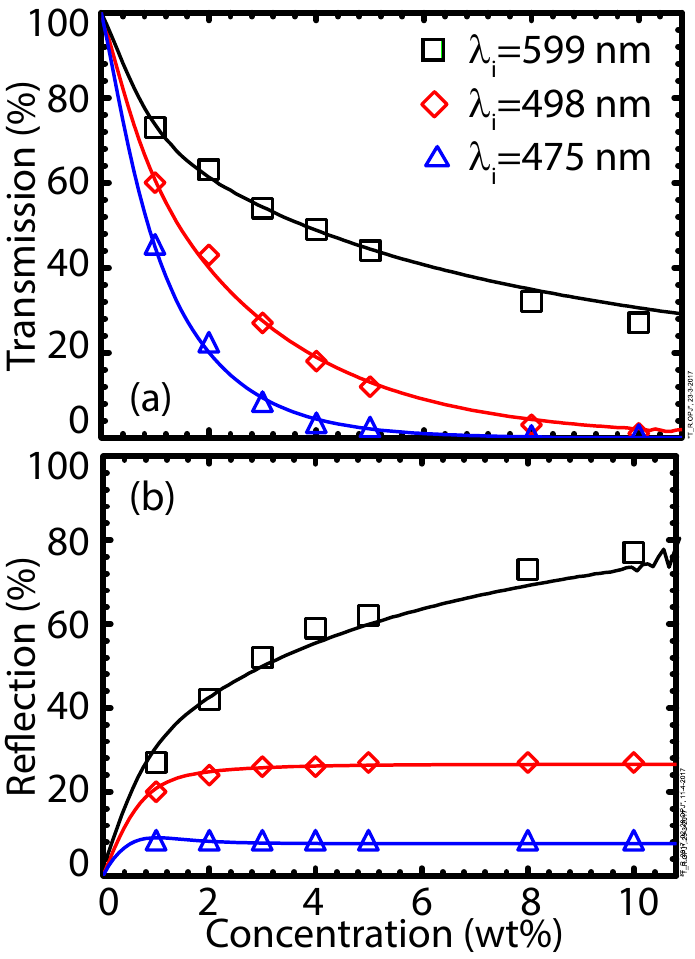}{3 in}{
Transmission and reflection as a function of concentration for three chosen wavelength.}{
\textbf{(a)} Transmission as a function of concentration shown for three different wavelength. \textbf{(b)} Reflection as a function of concentration shown for three different wavelength. On both figures symbols represent experimentally obtained values.
Solid lines are analytical results obtained using P3 approximation to the radiative transfer equation.}

We extract the transport parameters by applying the P3 approximation model Eq.~\ref{eq:f2a} and~\ref{eq:f2b} to the total diffuse transmission and the total diffuse reflection data. 
Since the theoretical description requires energy conservation with the diffuse reflection and transmission adding to $100~\%$, we decided to fit $T$ and $R$ separately in view of the slight systematic error mentioned above. 
As an example, the fit at $\lai=599$~nm is shown in Fig.~\ref{fig:Tw}. The difference between the anisotropy parameter $\mu (\lambda =599\mathrm{~nm})$ extracted from $T$ and $R$ is about 4~$\%$ percent, which is within the error bar of the experiment. 
Fig.~\ref{fig:Tw} illustrates the model prediction at three wavelengths in the range of strong absorption ($\lai$=475~nm), weak absorption ($\lai$=498~nm), and no absorption ($\lai$=599~nm).
The transmission is a monotonically decaying function of the concentration $\rho$ over the studied wavelength range. 
The diffuse reflection is an increasing function of the concentration $\rho$ in the wavelength range with no absorption. 
In the absorption range, the increase of the reflection $R$ in the high concentration range is hampered by the absorption. 

For each wavelength $\lai$ in the absorption range ($400<\lai<490$~nm), we fit the analytic ($T^{\mathrm{m}}$, $R^{\mathrm{m}}$) simultaneously to the measured (T,R) as a function of the concentration $\rho$, with the scattering cross section $\sss$, the absorption cross section $\sa$, and the anisotropy factor $\mu$ as the adjustable parameters. 
In the non-absorbing range $490~\mathrm{nm}<\lai<700~\mathrm{nm}$  the scattering cross sections $\sss$ is taken from the ballistic light measurements. 
For the stability of the procedure we take the absorption cross section to be small $\sa=10^{-7}$, where we verified that even a decrease of $\sa$ by several orders of magnitude does not affect the result.
The only adjustable parameter in the non-absorbing range is the anisotropy factor $\mu$. 
If the total transmission has strongly decreased to $T<1\%$ ,the analytical model reveal a strong oscillations shown in Fig.~\ref{fig:Tw}(b). 
These oscillations are the result of finite machine precision in calculating the numerical values of the total transmission $T$, and the total reflection $R$ using Eq.~\ref{eq:difsol}.

\FIGesmall{fig:s}{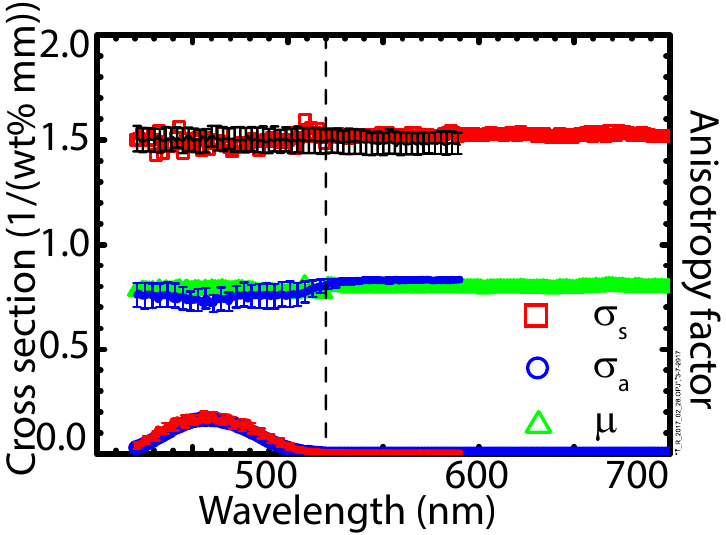}{3 in}{Transport parameters of $\Y$ phosphor particles extracted with analytical model and Monte-Carlo simulations.}{Open symbols are the transport parameters values extracted using analytical approach. The solid lines are the Monte-Carlo simulation results with respective error bars. Anisotropy factor $\mu$ ordinate is on the right hand side of the figure, the scattering cross section $\sss$ and the absorption cross $\sa$ section ordinate is on the left hand side of the figure. }

In Fig.~\ref{fig:s} the extracted parameters ($\sss$, $\sa$, $\mu$) are shown for the whole visible wavelength range. 
In the whole wavelength range the scattering cross section $\sss$, and the anisotropy factor $\mu$ remain constant with the incident wavelength $\lai$. 
In the absorption range the maximum absorption cross section $\sa$ coincides with the absorption peak of the $\Y$. 
The absorption cross section tends to zero at the edges of the absorption spectral range ($\lai=$420~nm and 520~nm) in agreement with Ref.~\cite{Meretska2016ContractorRept}. 

To verify the derivation of the parameters ($\sss$, $\sa$, $\mu$) by our analytical model, we employ Monte-Carlo simulations of the radiative transfer equation. At every wavelength $\lambda$, we simulate ($T^{\mathrm{s}}$, $R^{\mathrm{s}}$) as a function of ($\ls$, $\la$, $\mu$) to within 4~$\%$ of the measured (T,R). For a particular concentration $\rho$ this yields ($\ls$, $\la$, $\mu$) as a function of $\lambda$. We compute the parameter estimates for all measured concentrations $\rho$. We verify that ($\ls$, $\la$) are proportional to the density $\rho$, and that $\mu$ remains fixed for all densities $\rho$. The ($\sss$, $\sa$, $\mu$) at a given wavelength $\lambda$ is the best fit parameter set across all concentrations. We convert ($\ls$, $\la$) to ($\sss$, $\sa$) using Eq.~\ref{eq:pt1} and Eq.~\ref{eq:pt2}.

The simulation results reveal good agreement with the analytical results, as shown in Fig.~\ref{fig:s}, where the error bars in the Monte-Carlo parameter estimate is a result of the regression analysis. 
Hence the transport parameters of white LED can be extracted equally well with the P3 approximation to radiative transfer equation (RTE) and with the Monte-Carlo simulations.
We compared the sum of the scattering cross section $\sss$ and the absorption cross section $\sa$ with the sum obtained from the ballistic light measurements in the absorption range (see Fig.~\ref{fig:sssa}). Extracted transport parameters obtained with two different methods agree well with each other, which confirms the validity of our described model.

\section{Discussion}
In this paper, we have studied the transport of light through slabs that scatter and strongly absorb. 
We have extracted diffuse transport parameters for light ($\sss$, $\sa$, $\mu$) over the complete visible wavelength range using an analytical approach. 
We access the transport parameters in the strong absorption range $L/\la =4$ (see Fig.~\ref{fig:parameterspace}) that was previously not accessible for exact analytical description through the widely used diffusion theory. 
We verified that our analytical method is superior over numerical methods in speed and precision: 
It takes $\sim$1 min at each wavelength to obtain the transport parameters with errors $\Delta \sss/\sss=4\%$, $\Delta \sa/\sa=4\%$ and $\Delta \mu/\mu=4\%$ for analytical approach. The extraction speed of the analytical method can be further optimized using dedicated solvers. The precision of the analytical method within the much shorter processing time is only limited by the precision of the measured input data. To achieve comparable precision it takes $\sim$17 min for a Monte-Carlo approach with errors $\Delta \sss/\sss=4\%$, $\Delta \sa/\sa=10\%$ and $\Delta \mu/\mu=7\%$. The precision of the Monte-Carlo simulations scales as square root of the number of rays used.
Another advantage of the analytical approach over numerical methods is the predictive power over a broad range of transport parameters where $L/\la <7$. 
 
Let us place our approach in the context with previous work on the analytical extraction of transport parameters.
Leung \textit{et al.} reported the transport properties of $\Y$ plates using a filtered broadband light source, where the linear dependence of $\lt$ was exploited to calculate $\la$~\cite{Leung2014OE}. 
Firstly, the approximation used to analyze total transmission was employed outside its range of validity. 
Secondly, their described method is only valid in the region of strong absorption or emission, and not in the overlap region. 
Hence the described approach could not be employed to extract transport properties over the whole visible spectral range. 

Ref.~\cite{Meretska2016ContractorRept} reported the transport properties of $\Y$ plates using a narrowband light source that works in the whole visible spectral range including the overlap range ($490~\mathrm{nm}<\lai<520~\mathrm{nm}$). 
They calculated $\la$ under the assumption of the linear dependence of $\lt$ in the absorption range. 
The validity of this assumption was confirmed in this manuscript. 
Nevertheless, in the strongly absorbing region the employed diffusion theory is not valid, hence this method cannot fully capture the parameter range of white LEDs (see Fig.~\ref{fig:parameterspace}).

Gaonkar \textit{et al.}~\cite{Ganokar2014ApplOpt} used phenomenological theory (Kubelka-Munk theory) to decouple scattering and absorption properties of the diffusive samples. 
In the described approach the transport parameters are extracted using only an empirical relation between the Kubelka-Munk and the radiative transfer equation. 
Moreover, they extracted only the transport mean free path $\lt$ and the absorption mean free path $\la$, which gives no explicit information about the anisotropy factor $\mu$.

\section{Summary and outlook}
We have investigated the transport of light through slabs that scatter and strongly absorb, with a focus on white LEDs. 
We have extracted the transport parameters of white LED phosphor plates with in the visible wavelength range. 
We measured total transmission, total reflection, and ballistic transmission over the whole visible wavelength range. 
We analyzed the diffuse light measurements using the P3 approximation to the radiative transfer equation to extract the scattering cross-section $\sss$, the absorption cross-section $\sa$ and the anisotropy factor $\mu$, in scattering media with strong absorption, up to $L/\la=4$. 
The Lambert-Beer law was used to analyze the ballistic light measurements, and extract $\sss+\sa$ in the whole visible spectral range. 
We verified the data obtained with the diffuse light measurements to the data measured with the ballistic light, and Monte-Carlo simulations. 
Our study confirms the validity of the P3 approximation for light propagation in medium suitable for white LEDs (see Fig.~\ref{fig:parameterspace}).

The procedure of finding transport parameters is a main hindering step for the conventional light propagation modeling, such as ray tracing~\cite{Gilray1996IESNA,Cassarly2001Book,Sommer2009JSTQE,Liu2010AO,Tukker2010SPIE}. 
Our method avoids the main bottleneck of these conventional numerical methods, providing $17$-fold speed up in time with the same error. 
Thus, our approach is a suitable analytical tool for industry, since it provides a fast yet accurate determination of key transport parameters, and more importantly it introduces predictive power into the design process of white LEDs. 

\section{Acknowledgments}
It is a great pleasure to thank Jan Jansen for sample fabrication, Cornelis Harteveld for technical support, and Teus Tukker, Shakeeb Bin Hasan, Oluwafemi Ojambati, Diana Grishina for useful discussions. 
This work was supported by the Dutch Technology Foundation STW (contract no. 11985), and by the FOM program "Stirring of Light!", by the Dutch Funding Agency NWO, and by the ERC.

\appendix
\section{Appendix: The P3 approximation to the radiative transfer equation (RTE)}
In this Appendix we briefly sketch the radiative transfer equation and its solution. 
For extensive background, we refer to Refs.~~\cite{Star1989SPIE,Boas1995ProcSPIE,Dickey1998PhysMedBiol,Dickey2001PhysMedBiol,Faris2005ApplOpt,Klose2006JComputPhys,Liemert2014MedPhys}. 
In the slab geometry the radiative transfer equation for light is 
\begin{equation}\begin{split} 
\nonumber
\mu \frac{\partial \Izmud}{\partial z} =& -\rho \st \Izmud \\
&+\rho \st \int_{-1}^{1}\pmm\Izmupd d\mu^\prime \nonumber\\
&+\frac{\rho \sigma_{\mathrm{t}}}{4\pi}\int_{4\pi}\pss \Ibp d\omega^\prime \mathrm{,}
\end{split} 
\label{eq:f3}
\end{equation}
where $\pmm$ is the phase function, and all other parameters and variables are given in section~\ref{sect:theory_meas}.
We expand the intensity $\Idd$ in spherical harmonics~\cite{Liemert2014MedPhys}
\begin{equation}
\Idd=\sum\limits_{l=0}^{N} \sum\limits_{m=-l}^{l} \flm \Ylm
\mathrm{,}
\label{eq:f1}
\end{equation}
that we substitute in Eq.~(\ref{eq:f3}). 
The properties of the Legendre polynomials allow us to simplify equation~(\ref{eq:f3}) to:
\begin{eqnarray}\nonumber
(2m+1) \rho\st \exp(-\rho\st z)\mathrm{W}_{\mathrm{m}}=&(m+1)\frac{d \fmpz}{d z}+m \frac{d \fmmz}{d z}\\
&+\rho \st \left(1-\mathrm{W}_{\mathrm{m}}\right)(2m+1)\fm
\mathrm{,}
\label{eq:exp1}
\end{eqnarray}
where $\mathrm{W}_{\mathrm{m}}$ are the moments of $\pmm$, and $m=0,1,2,3$. 
The set of equations~(\ref{eq:exp1}) has particular and complementary solutions that are equal to
\begin{equation} 
\psi_m=\sum\limits_{i=1}^{2}H_{\mathrm{mi}}C_\mathrm{i} \exp{(\mu_i z)}+G_\mathrm{m} \exp{(-\rho\st z)}
 \mathrm{,}
\label{eq:difsol}
\end{equation}
where $m=0,1,2,3$, $i=1,2,3,4$, $H_{\mathrm{mi}}$, $C_\mathrm{i}$, and $G_\mathrm{m}$ are the coefficients obtained by the substitution of the Eq.~\ref{eq:difsol} in the Eq.~\ref{eq:exp1} and in the boundary conditions. 
Finally, the light intensity is obtained when Eq.~(\ref{eq:difsol}) is substituted into Eq.~(\ref{eq:f1}).

\end{document}